# An adaptive random experiment design method for engineering experiment

Zhou Qiao[1], Duan Xiaochang[2], Tang Wei[2, *]

1. School of Environment and Safety Engineering, North University of China, Taiyuan 030051, China.

2. Institute of Chemical Materials, China Academy of Engineering Physics, Mianyang 621999, China.

**Abstract**

This paper proposes an adaptive random experiment design (ARED) algorithm that can be applied to optimize the multiple factors and levels experiments. The algorithm takes real-time model error as the adaptive condition, and outputs a model that conforms to the error quantization standard based on the automatic process. According to the actual experimental scenario, the similar number of test cases were selected between the ARED method and the comparative experimental design method under the bimodal Gaussian function, the bimodal surface function and the peaks function, respectively. simultaneously, the support vector machine (SVM) algorithm is used to construct the model for the selected test cases, and the verification surface (or curve) is predicted. The qualitative and quantitative analysis is carried out at two-slice of applicability and precision. The results show that the ARED method can be applied to the experiment of multi-factor, and has better precision and applicability than the comparative experimental methods.

**1 Introduction**

The design of experiment (DOE) is a branch of mathematical statistics, which is a mathematical principle and implementation method for formulating an appropriate experimental plan, and performing effective statistical analysis according to a predetermined target [1]. Uniform experimental design [2], orthogonal experimental design [3], factorial experimental design [4] and other DOE methods have solved many engineering [5], education [6] and scientific [7] problems for human beings.

The selection of appropriate factors and corresponding levels in the experimental design is a key of the specific procedures and data analysis for the experiment [8]. Currently, most of the factors and levels are artificially defined, and the DOE method requires a certain correspondence between the number of factors and levels [9]. However, in engineering, the value of the levels may be infinite within a certain factor (such as temperature, pressure, etc.), and it will be created the irrationality numbers of factors and levels for human being under the unknown law of experiment. Scilicet, we cannot distribute the more factors and levels for more intricate law of experiment under the unknown law of experiment. Moreover, In the conventional DOE methods, the factors and levels are selected first, then the experimental test is carried out, and finally the linear process of mathematical model is established [10-11] by regression analysis [12] or response surface method (RSM) [13]. However, these methods do not take into account the accuracy of the output model. If the output model is difficult to meet the applicable standards, it may be necessary to further supplement the experiment or re-design the experiment, resulting in waste of resources. In summary, it is necessary to design an experiment design method that can cope a variety of factors and levels, and output a model with quantitative criteria.

Therefore, the adaptive random testing (ART) method is proposed to solve these problems. ART technology is a subfield of software engineering [14]. It is a method of randomly drawing test cases under certain rules to test software defects [15]. Due to its simplicity and high efficiency, adaptive thinking has been widely used in algorithm optimization [16], computerized adaptive testing [17], industrial processes design [18] and other fields.

In this paper, an adaptive random experiment design (ARED) algorithm is compiled that can cope manifold factors experiment. It includes selecting initial samples, constructing constraints, drawing test cases, establishing support vector machine (SVM) models, analyzing the model error, determining the error feedback, and outputting the quantification models. In addition, comparing with the single factor experiment (SFE) method and factorial experiment method, the qualitative and quantitative results under the bimodal Gaussian curve function, the bimodal surface function and the peaks function will obtain.

**2 Adaptive random experiment design method**

The main components of the ARED algorithm include: determining the sample set dimension, selecting the initial sample, drawing the test case under the constraint condition, establishing the SVM model, error analysis, error feedback and model output. The blanket flowchart is shown in Fig. 1.

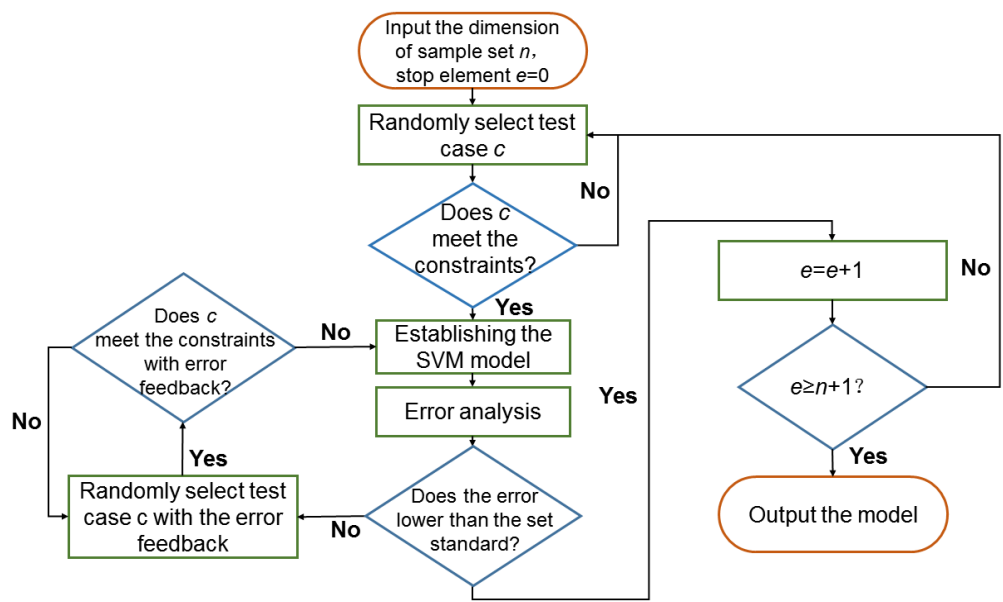

Fig.1 The flowchart for adaptive random experiment design.

**2.1 Determining the sample set dimension**

The dimensions of the sample set originate the number of experimental variable. Taking the creep-tension essay in mechanics as an example, the strain-time curve is obtained under the uniaxial quasi-static creep tensile test. The strain is independent variable (iv) and the time is dependent variable (dv). At this point, the sample set has two dimensions. If the temperature change is added, the sample set becomes 3D, and considering the situation of the confining pressure, it adds to 4D.

In the n-dimensional experiment, the input domain is $\mathbf{A}=[\mathbf{a_1}, \mathbf{a_2}, \mathbf{a_3} ... \mathbf{a_n}]$, which contains **n** variables (vectors). Each variable consists of $m_i$ values (scalars), which are $[a_{11}, a_{12}, a_{13} ... a_{1m1}]$.

**2.2 Selecting the initial sample**

The initial samples are the experimental condition need to be performed before the beginning of the algorithm, which contains at least the start and end points of each variable (e. g. $[a_{11}, a_{1m1}]$). In addition, the known key points also can be added, including but not limited to inflection points and extreme points (e.g. glass transition temperature ($T_g$) point). In this algorithm, it plays a decisive role in controlling the domain of variable.

**2.3 drawing the test case**

Test cases are selected by an ARED algorithm to guide the experimental process. In the n-dimensional experiment, the test case is a random variable **c** whose dimension is the same as the sample set, and its value from the random value of the per experimental variable.

For example, $\mathbf{c_1}= [a_{1k}, a_{2k}, a_{3k} ... a_{nk}]$, where k= $u$ (x; $b_1$, $b_2$), $u(x)$ is a probability density function, $b_1$, $b_2$ are the range of $u(x)$, and $b_2>b_1$. In order to ensure that the key points can be find by test cases, the selection principles are as follows:

(1). The coordinates of all the test cases need to be within the domain of definition.

(2). The algorithm should be avoiding the influence of human factors.

(3). Extreme points have a large impact on the trend of the region. The worth of the test cases at this location is higher, which increases the probability that the test cases will be selected in this area. When the law of experiment is unknown, the mid-value of the variables can be used as the high value region (this position is the furthest from the sum of all initial sample distances).

(4). In the algorithm, the test cases are not expectantly selected near the initial samples, because the worth of cases are lower around this region.

(5). The selected test cases should be sufficiently scattered, because it has only one case reference value when the multiple test cases gathered together.

**2.4 constrained condition**

**2.4.1 Random function**

In order to meet the above conditions, several constraints must be added to satisfy. Considering that the probability density curve of a normal distribution conforms to the selection design principle, and the experimental error is also subject to the normal distribution [19, 20]. In consequence, the probability density distribution function adopts the normal distribution, such as Eq. (1).

$$f(x) = \frac{1}{\sqrt{2\pi}\sigma} e^{-\frac{(x-\mu)^2}{2\sigma^2}}, -\infty < x < \infty \quad (1)$$

Where the $\mu$ is expectation, $\sigma$ is variance.

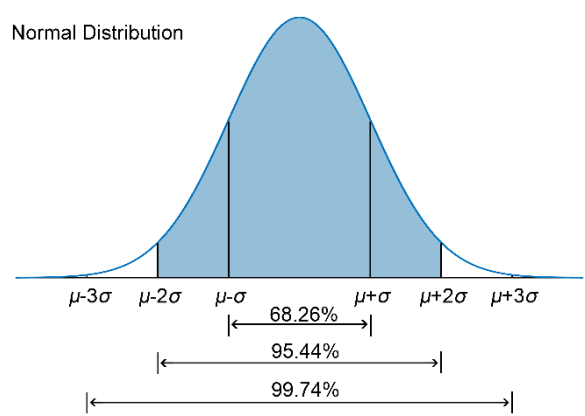

Fig.2 Abridged general view of normal distribution

A schematic of normal distribution is shown in Fig.2, the region of input domain is $\mu \pm 2\sigma$ (the confidence interval is 95%), which is indicated by the blue region in Fig. 2. This method can reduce the probability of a test case appearing near the initial sample.

In addition, to avoid the multiple test cases gathered together, the constraint of minimum distance between the two test cases is proposed. The illustration of 2D and 3D sample set as follows, and the higher dimension issue can be generalized on this basis.

**2.5.1 Two dimension sample set**

The 2D sample set has one iv **I**, and its domain is [0, i]. In Figure 3, the blue "●" represent the initial samples, the green "●" represent the selected test cases, and the red "●" represents the test case selected this round.

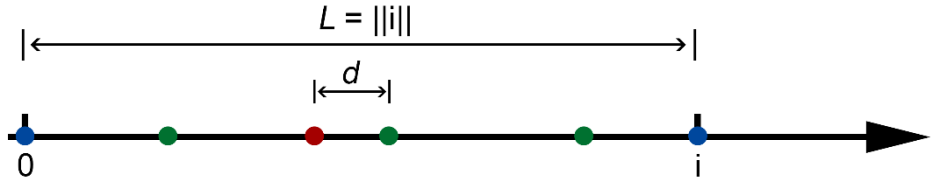

Fig.3 The constrained condition of 2-dimension sample set

The condition for this selected test case is that the distance d (the distance between the test case selected this round and the nearest case in this number axis) is larger than the ratio of the domain length L to the control function (p· v + q). In addition, as the number of selected test cases increases, the control function is defined as a decreasing function to avoid the algorithm enter an infinite loop.

The constrained condition as shown in Eq. (2):

$$d > \frac{L}{pv + q} \tag{2}$$

Where the $v$ is the number of test cases. $p$ and $q$ are control coefficients, which can be adjusted according to requirements. According to our experience, for a single independent experiment, $p$ is 0.7 and $q$ is 10. If the test case selected this round does not meet the constraint, re-run the loop until it meets the limits of Eq. (2).

**2.5.2 Three dimension sample set**

For the experiment of double ivs, the constraint conditions are similar to the 2D sample set. The ivs of the sample set are setting to **I** and **J**, whose domain are [0, i] and [0, j], respectively. The blue "●", green "●", and red "●" have the same meaning as above. The schematic diagram is shown in Fig. 4.

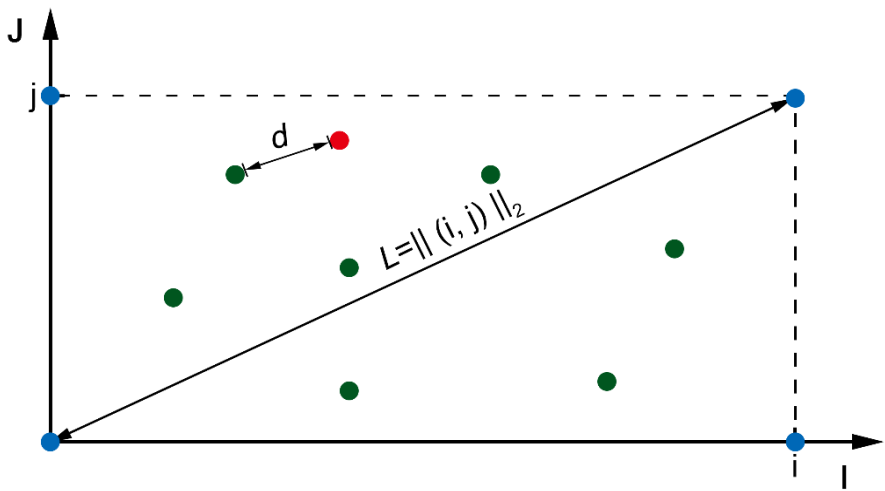

Fig.4 The constrained condition of 3-dimension sample set

Where the constraint L becomes the furthest distance in the region. The $p$ and $q$ can be defined as

0.4 and 5, respectively. For higher-dimensional experiments, it is only necessary to extend the description of d and L over the Euclidean distance. The calculation method and formula remain unchanged.

**2.6 Established the SVM model**

After a qualified test case is obtained, all the selected cases and initial samples is used as training set, the value of the test case selected this round is predicted by SVM algorithm.

The support vector machine was first proposed by Vapnik and Chervonenkis in 1964 [21], through building a hyperplane between two sets train samples. The linear separability problem was originally solved by the algorithm up to the time when Smola and Scolkopfs [22] applied SVM to the solution of the regression problem. SVM algorithm is a proverbially method of a computer science subfield that provides solutions in the fields of failure surface modeling [23], computational acceleration [24], data classification [25], and so on.

In this paper, the kernel function of the SVM algorithm is radial basis function (RBF) [26]. For the RBF kernel function, it has two parameters of penalty parameter C and the kernel parameter γ. In order to obtain the optimal parameters combination, the optimization strategy need to be applied. Considering that the parameters are different after each training, the fine-grained grid-search algorithm and the K-fold cross-validation (K-CV) algorithm are used to perform the parameter optimization process. The search range of C and γ is $10^{-11}$ to $10^{11}$, and the search step is 0.5.

The training set is incremented after each selection, and new models need to be constantly trained to accommodate changes in the training set. Therefore, the incremental learning [27] is introduced to create model after each test case is selected, and the predicted value of the test case selected this round is obtained.

**2.7 Error analysis**

Restoring the test case to experiment conditions, experimentizing, and entering the actual value. The absolute percent error (APE) and absolute error (Abs E) of the predicted and actual values at all cases is calculated. Related calculation formulas are Eq. (3), Eq. (4).

$$E_{ape} = \frac{1}{n} \times \sum_{i}^{n} \left| \frac{(x_i - y_i)}{y_i} \right| \times 100\% \tag{3}$$

$$E_{ae} = \frac{1}{n} \times \sum_{i}^{n} |x_i - y_i| \tag{4}$$

Where n was the quantity of globe samples, x was the theoretical value, y was the actual value.

When the model error meets the set standard, the ARED algorithm continues to proceed. Otherwise, enter the error feedback and select the test case.

**2.8 Error feedback**

**2.8.1 The condition of error feedback**

The error feedback is required when the error of samples in accordance with the three and relationships.

(1). To ensure that the existing selected cases can be formed a basic model, the error feedback is enabled when the number of cases is greater than $n^2+1$;

(2). In the error analysis, the APE of a case exceeds $E_{ape}$%;

(3). The Abs E of this case exceeds $r$% of the known largest range (the value of $r$ can be selected according to the material dispersibility, which is taken as 10% in this paper).

When the three conditions are met at the same time, the constraint condition is changed to increase the probability that the next case appears within the range of the maximum error case. In order to achieve it, all the cases are marked that meet the feedback conditions at first, and then sort them based on the relative error. Finally, the test case with the largest relative error is taken as the feedback point.

Taking the iv value feedback case as the expected $\mu$, the 10% of the domain length is defined as the variance $\sigma$, the $\mu\pm2\sigma$ is the domain of the error feedback, and the test case is taken under the Eq. (1). Let the coordinate of the feedback case be $(x_t, y_t)$, then the schematic diagram of error feedback is shown in Fig. 5.

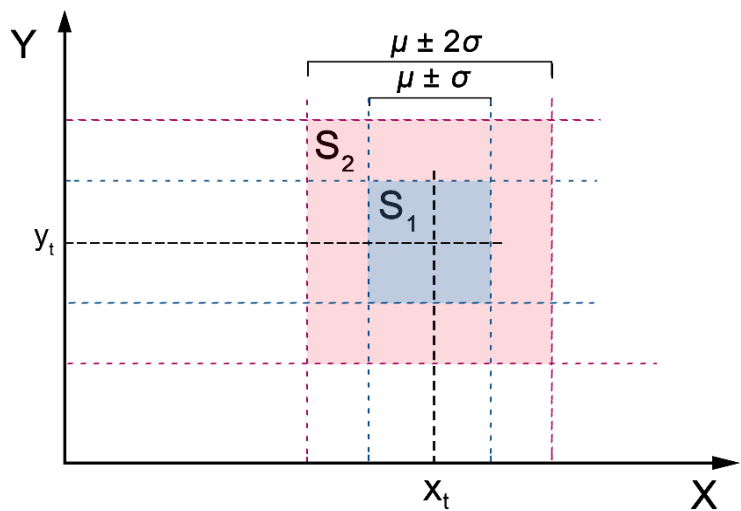

Fig. 5 Schematic diagram of 3D sample set error feedback. The blue region ($S_1$) refers to the $\mu\pm\sigma$ interval, which is

20% of the domain length and accounts for 4% of the total region. The red region ($S_2$) refers to the $\mu \pm 2\sigma$ interval, which is 40% of the domain length and accounts for 16% of the total region.

**2.5.2 The constraint conditions of error feedback**

The test cases will inevitably accumulate when using error feedback to select cases. Therefore, the new constraints are formulated for error feedback.

(1). The coordinates of the feedback test cases need to be within the red region ($S_2$);

(2). The experiment should be avoiding the influence of human factors.

Under the new constraints, the formula is still the Eq. (2). For the value of the control variable, p is 1.5 and q is 15 in the 2D experiment. The 3D experiment can take p as 7 and q as 0.5.

**2.9 Model Export**

The experiment is continued until the SVM model error of the continuous $n+1$ times is lower than the error feedback standard, and the model training is completed and outputted.

**3. Adaptive random testing method test**

In order to verify the effectiveness of this algorithm, the algorithm was tested in 2 and 3 dimensional experiments. To ensure the reliability of the results, the existing functions were used for testing.

**3.1 Two dimension experiment test**

**3.1.1 Bimodal Gaussian function**

In the 2D experiment, the test curve was assumed to be a bimodal Gaussian function with extreme positions of 0.3126, 0.456, and 0.6758, respectively, and the curve consists of 10,000 scatter points. The function forms were shown in the Eq. (6), Eq. (7), Eq. (8), Eq. (9) and Eq. (10), and the geometric form of the function was shown in Fig. 6.

$$c_1 = \frac{A_1 \times SIG_1}{\sqrt{2\pi}} \tag{6}$$

$$k_1 = 2 \times SIG_1^2 \tag{7}$$

$$c_2 = \frac{A_2 \times SIG_2}{\sqrt{2\pi}} \tag{8}$$

$$k_2 = 2 \times SIG_2^2 \tag{9}$$

$$p = K + c_1 \times e^{-\frac{(z-M_1)^2}{k_1}} + c_2 \times e^{-\frac{(z-M_2)^2}{k_2}} \times 10000 \tag{10}$$

Where the $A_1$ and $A_2$ were the peaks of the formula, and the values were 0.2 and 0.3, respectively. The respective expectations for the peaks were $M_1$ and $M_2$, with values of 0.3126 and 0.6758, respectively.

The $SIG_1$ and $SIG_2$ were corresponding standard deviations, and the values were 0.1 and 0.2, respectively. K was the baseline with the value of 0.0002. z was the number of scatters, which was 10,000.

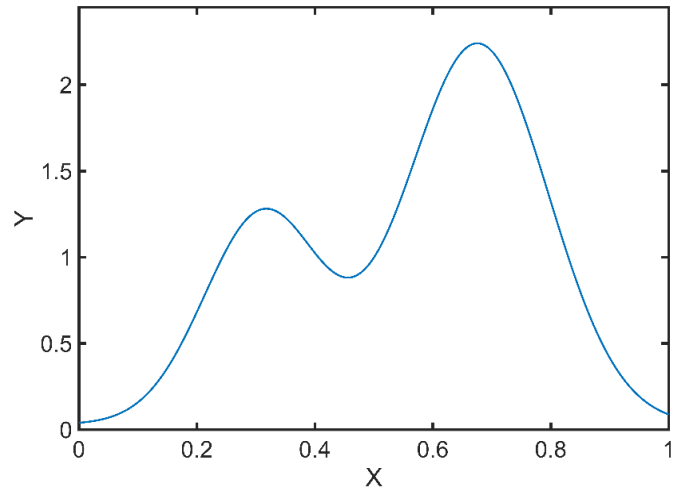

Fig.6 Geometric form of Bimodal Gaussian function

**3.1.2 Algorithm test of bimodal Gaussian function**

(0, 0.0401) and (1, 0.0882) were the initial samples, and the test case was selected under the constraint conditions until the end of the algorithm. Let the algorithm run 10 times in a row to collect data. During the test, the test cases were selected up to 14 times, at least 8 times with an average of 12.7 times. At the same time, the same number of test cases were taken equidistantly within the domain (in Fig. 7). As can be seen from Fig. 7, both methods could select test cases around 0.31, 0.45, and 0.67. However, in Test-5, the simple factor experiment (SFE) did not produce test cases around 0.45.

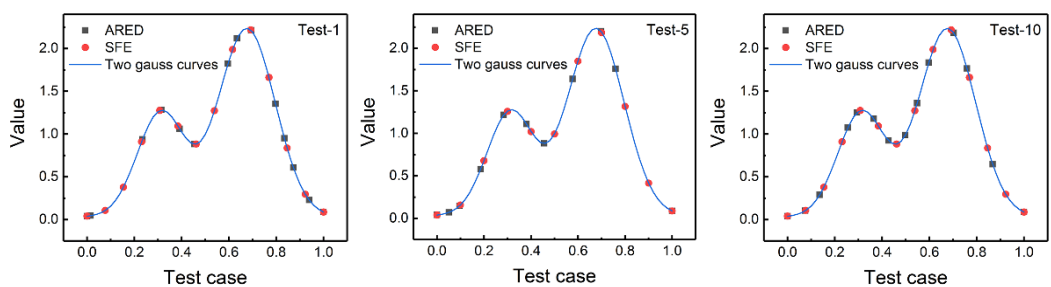

Fig. 7 The selected test cases of ARED and SFE in bimodal Gaussian function

In addition, the SVM algorithm was used to model the two sets of sample, and the verification curve with 50 samples (in Fig. 8) was predicted by this model. For Fig. 8, the ARED method could accurately predict the bimodal Gaussian curve in these two cases. However, in Test-2, the selected test cases according to the SFE could not describe the function correctly. It could be deemed that the ARED method was more applicable than the SFE method.

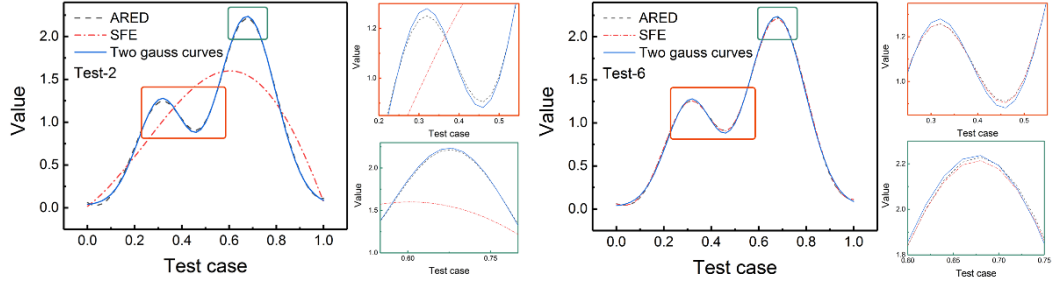

Fig. 8 The prediction results of ARED and SFE in bimodal Gaussian function

**3.1.3 Error analysis of bimodal Gaussian function**

In order to quantify the accuracy, the model of ARED and SFE was compared with the verification curve, the mean absolute error (MAE), mean absolute percentage error (MAPE) and coefficient correlation (R) was calculated simultaneously. The calculation formulas of MAE, MAPE, and R were shown in Eq. (11), (12), and (13).

$$MAPE = \frac{1}{n} \cdot \sum_{i}^{n} \left| (x_i - y_i) / y_i \right| \times 100 \tag{11}$$

$$MAE = \frac{1}{n} \cdot \sum_{i}^{n} \left| x_i - y_i \right| \tag{12}$$

$$R = \frac{Cov(x, y)}{\sqrt{D(x)}\sqrt{D(y)}} \tag{13}$$

Where n was the quantity of globe samples, x was the theoretical value, y was the actual value.

Table 1 presented the results of the model error. It could be found that the ARED method maintained high precision at wholly tests, and the SFE method was also held the high precision expect the Test-2 and Test-3 (the number of samples was 12).

Among them, the maximum MAE, MAPE and the minimum R of the ARED method were 0.03596, 6.995% and 0.9981, respectively, which appeared in Test-4. The maximum MAE, MAPE and minimum R of the SFE were found in Test-2 and Test-3, and the values were 0.2602, 37.84% and 0.9174, respectively. If the Test-2 and Test-3 were not calculated, the maximum MAE and MAPE was 0.0469 and 9.235%, respectively, and the minimum R was 0.9976. Therefore, the ARED method had more applicability in the 2D sample set than the SFE method.

However, it should be noted that when the number of cases exceeded 13, the accuracy of the SFE was significantly improved. For example, in Test-8, the quantitative standard of SFE was better than the

ARED. And when the number of samples was less than 14, the accuracy of the ARED method is better than the SFE method. Therefore, it can be considered that as the number of samples increases, the advantages of ARED were gradually reduced, and eventually inferior to SFE method. Taking adaptive meshing in finite element software [28] as an example, adaptive meshing was not necessary when the mesh was sufficiently fine.

Table 1 Errors analysis of ARED method and SFE method under 2D experiment

| No. | The number of sample | Source | MAE | MAPE (%) | R |
|---|---|---|---|---|---|
| 1 | 14 | ARED-1 | 0.01657 | 4.742 | 0.99971 |
| 2 |  | SFE-1 | 0.01669 | 5.051 | 0.99973 |
| 3 | 12 | ARED-2 | 0.01654 | 6.312 | 0.9997 |
| 4 |  | SFE-2 | 0.2602 | 37.84 | 0.9174 |
| 5 | 12 | ARED-3 | 0.01833 | 5.997 | 0.9997 |
| 6 |  | SFE-3 | 0.2602 | 37.84 | 0.9174 |
| 7 | 10 | ARED-4 | 0.03596 | 6.995 | 0.9981 |
| 8 |  | SFE-4 | 0.04691 | 9.235 | 0.9976 |
| 9 | 11 | ARED-5 | 0.01627 | 5.519 | 0.9997 |
| 10 |  | SFE-5 | 0.02251 | 6.738 | 0.9995 |
| 11 | 14 | ARED-6 | 0.01627 | 5.519 | 0.99974 |
| 12 |  | SFE-6 | 0.01669 | 5.051 | 0.99973 |
| 13 | 11 | ARED-7 | 0.02186 | 5.396 | 0.9996 |
| 14 |  | SFE-7 | 0.02251 | 6.738 | 0.9995 |
| 15 | 16 | ARED-8 | 0.01760 | 5.707 | 0.9997 |
| 16 |  | SFE-8 | 0.01554 | 4.966 | 0.9998 |
| 17 | 13 | ARED-9 | 0.01680 | 5.636 | 0.9997 |
| 18 |  | SFE-9 | 0.02107 | 5.713 | 0.9995 |
| 19 | 14 | ARED-10 | 0.02241 | 6.568 | 0.9995 |
| 20 |  | SFE-10 | 0.01669 | 5.051 | 0.9997 |

**3.2 Three dimension experiment test**

**3.2.1 Function of surface**

For the test of 3D sample set, the bimodal surface function was purposed to verify the accuracy and applicability of the ARED method. The function form was shown in Eq. (14), and its geometric form was shown in Fig. 9.

$$z = y \times e^{(-x^2 - y^2)} \times 50 \tag{14}$$

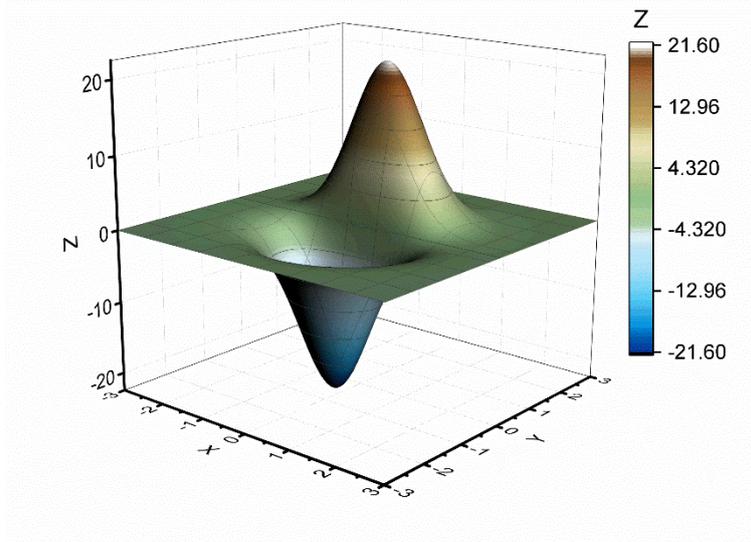

Fig. 9 Geometric form of bimodal surface function

**3.2.2 Algorithm test of bimodal surface function**

Let the algorithm run 10 times in a row to collect data, the detailed data were shown in Fig. S. 2. The initial samples were (-3, -3, -2.28E-06), (-3, 3, 2.28E-06), (3, -3, 1.37E-05) and (3, 3, 2.28E-06), and the test cases were selected from the ARED. During the trial period, the test cases were selected up to 37 times, at least 21 times were selected, with an average of 32.4 times. At the same time, the factorial experiment (FE) was used to select test cases with a similar or more number (If the ARED method selected 36 cases, the FE selected 36 cases. If the ARED method selected 33 cases, the FE selected 36 cases.). In Fig. 10, compared with the FE, the test points selected by the ARED were closer to the maximum point, and the two methods were closed to the minimum point.

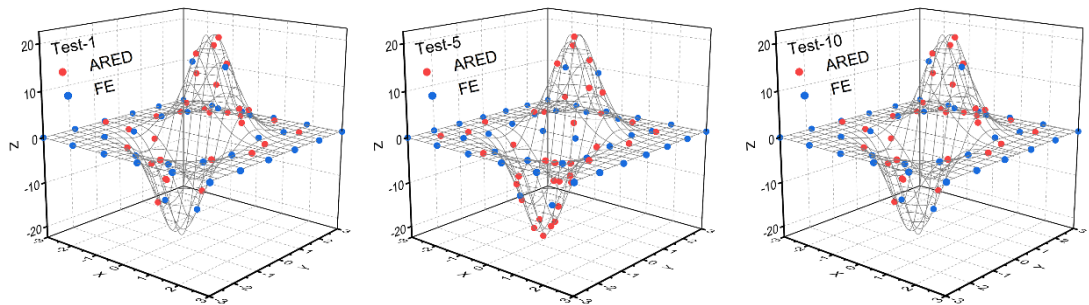

Fig.10 The selected test cases of ARED and FE in bimodal surface function.

Similarly, the SVM algorithm was used to model the two sets of data and to predict a verification surface with 121 data points (Fig. 11). As shown in Fig. 11 a, the test surface can be described correctly via test cases selected by the FE method and the ARED method, and the accuracy of the ARED method was higher than FE method. In addition, when the number of cases was 25 and 30 (corresponding to Test-

6, Test-8, Test-9 of the FE, Fig. 11 b.), the FE method had also appeared in the 2D experimental test, which did not correctly describe the verification curve.

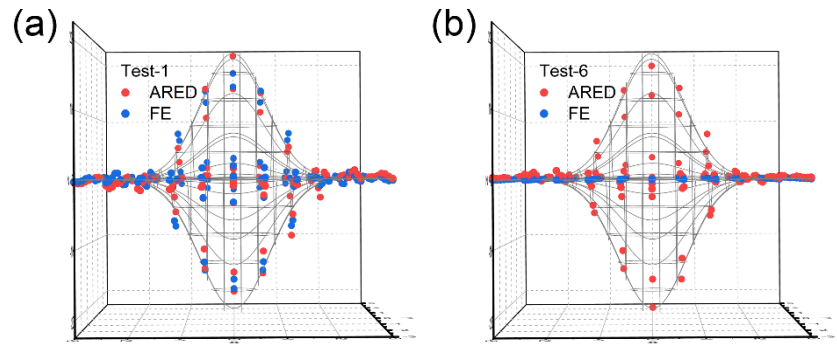

Fig. 11 The prediction results of ARED and FE in bimodal surface function.

**3.2.3 Error analysis of bimodal surface function**

In error analysis, since iv Y=0, regardless of the value of iv X, the dv Z value was 0, but the predicted value may not be equal to 0. Therefore, the value of MAPE becomes infinity at this point, but the MAE kept a lower level.

For Table 2, the MAPE was discarded. The maximum MAE and minimum R of the ARED method appeared at Test-8, which were 0.8733 and 0.9626, respectively. The maximum MAE and minimum R of FE appeared at Test-6 and Test-9, which was 1.906 and 0.4932, respectively. From the results, the advantage of the ARED method was highlighted when the experimental dimension was increased, which exceeds the FE method in both accuracy and applicability.

Table 2 Errors analysis of ARED method and FE method under 3D experiment

| No. | The number of sample | Source | MAE | R |
| --- | --- | --- | --- | --- |
| 1 | 35 | ARED-1 | 0.5825 | 0.9879 |
| 2 | 36 | FE-1 | 0.7143 | 0.9787 |
| 3 | 33 | ARED-2 | 0.3944 | 0.9939 |
| 4 | 36 | FE-2 | 0.7143 | 0.9787 |
| 5 | 32 | ARED-3 | 0.5054 | 0.9905 |
| 6 | 36 | FE-3 | 0.7143 | 0.9787 |
| 7 | 33 | ARED-4 | 0.3151 | 0.9945 |
| 8 | 36 | FE-4 | 0.7143 | 0.9787 |
| 9 | 41 | ARED-5 | 0.2030 | 0.9983 |
| 10 | 42 | FE-5 | 0.7837 | 0.9658 |
| 11 | 28 | ARED-6 | 0.5085 | 0.9923 |
| 12 | 30 | FE-6 | 1.9060 | 0.4932 |
| 13 | 34 | ARED-7 | 0.5338 | 0.9913 |
| 14 | 36 | FE-7 | 0.7143 | 0.9787 |
| 15 | 25 | ARED-8 | 0.8733 | 0.9626 |
| 16 | 25 | FE-8 | 1.7909 | 0.5723 |
| 17 | 28 | ARED-9 | 0.5963 | 0.9722 |
| 18 | 30 | FE-9 | 1.9060 | 0.4932 |
| 19 | 35 | ARED-10 | 0.5824 | 0.9879 |
| 20 | 36 | FE-10 | 0.7143 | 0.9787 |

**3.3 Complex surface test**

**3.3.1 Peaks function**

The peaks function was used as a probability density function for testing the binary Gaussian distribution of 3D plots. It had 3 local maximums and 3 local minimums, as shown in Fig. 12, and its function form was shown in Eq. (15). In this chapter, the characteristics of the ARED method were further explored on the theoretical level.

$$z = 3(1-x)^2 \times e^{\left(-x^2-(y+1)^2\right)} - 10\left(\frac{x}{5} - x^3 - y^5\right) \times e^{\left(-x^2-y^2\right)} - \frac{1}{3} e^{\left(-(x+1)^2-y^2\right)} \tag{15}$$

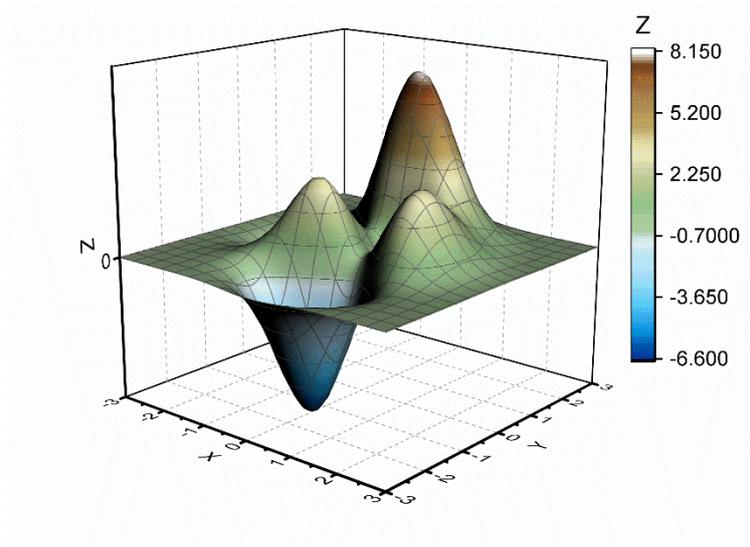

Fig.12 Geometric form of peaks function

**3.3.2 Algorithm test of peaks function**

Uniformly, the points of (-3, -3, 6.67E-05), (-3, 3, 3.22E-05), (3, -3, - 5.86E-06) and (3, 3, 4.1E-05) were selected as the initial sample, and the ARED algorithm was continuously run 10 times to collect data. During the test duration, the test cases were selected up to 50 times, and at least 30 times were selected, with an average of 35 times. At the same time, the FE method was used to select test cases. Part test results were shown in Fig. 13.

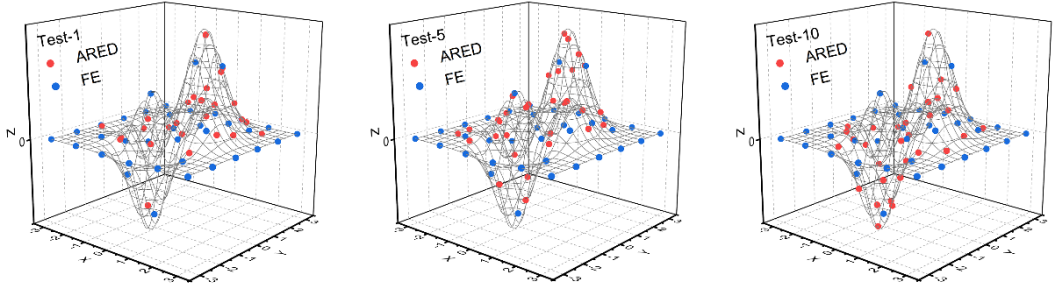

Fig. 13 The selected test cases of ARED and FE in peaks function.

Built the model and predicted the verification surface, the representative results were shown in Fig. 14. When the number of sample points was less than 42 (Test-1, Test-4, Test-5, Test-10), the ARED method could describe the verification surface correctly (Fig. 14 a). However, the model cannot be correctly established by FE, which was difficult to describe the peaks function. For Fig. 14 b, the number of test cases was greater than the 42, the accuracy of the FE method had rebounded, but it was still inferior to the ARED method.

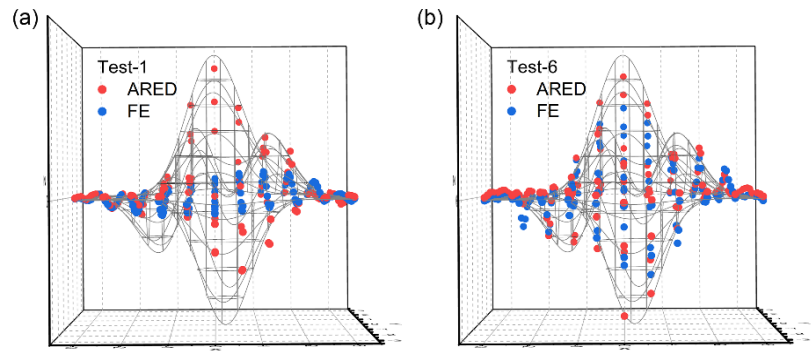

Fig. 14 The prediction results of ARED and FE in peaks function.

### 3.3.3 Error analysis of peaks function

The detailed error calculation results were shown in Table 3. the MAE value of the ARED method was 0.007 higher than the FE (only in Test-9), and others MAE values were lower than the FE. The maximum MAE and minimum R of the ARED method were found in Test-9, which values were 0.4988 and 0.9639, respectively. The maximum MAE and minimum R of FE appeared when the number of selected cases was 36, and the values were 0.7864 and 0.8916, respectively. Therefore, when the complexity of the experimental surface was increased, the accuracy and applicability of the ARED method were more advantageous.

Table 3 Errors analysis of ARED method and SFE method under peaks function

| No. | The number of sample | Source | MAE | R |
| --- | --- | --- | --- | --- |

| 1  | 35 | ARED-1  | 0.4675 | 0.9702 |
| 2  | 36 | FE-1    | 0.7864 | 0.8916 |
| 3  | 33 | ARED-2  | 0.2299 | 0.9967 |
| 4  | 36 | FE-2    | 0.4918 | 0.9703 |
| 5  | 32 | ARED-3  | 0.3155 | 0.9916 |
| 6  | 36 | FE-3    | 0.4918 | 0.9703 |
| 7  | 33 | ARED-4  | 0.4346 | 0.9720 |
| 8  | 36 | FE-4    | 0.7864 | 0.8916 |
| 9  | 41 | ARED-5  | 0.3130 | 0.9873 |
| 10 | 42 | FE-5    | 0.7864 | 0.8916 |
| 11 | 28 | ARED-6  | 0.2747 | 0.9944 |
| 12 | 30 | FE-6    | 0.4918 | 0.9703 |
| 13 | 34 | ARED-7  | 0.3277 | 0.9780 |
| 14 | 36 | FE-7    | 0.4918 | 0.9703 |
| 15 | 25 | ARED-8  | 0.2337 | 0.9951 |
| 16 | 25 | FE-8    | 0.4846 | 0.9705 |
| 17 | 28 | ARED-9  | 0.4988 | 0.9639 |
| 18 | 30 | FE-9    | 0.4918 | 0.9703 |
| 19 | 35 | ARED-10 | 0.3970 | 0.9719 |
| 20 | 36 | FE-10   | 0.7864 | 0.8916 |

## 4. Conclusion

Based on the adaptive thinking, this paper constructs an adaptive random experiment design algorithm. The algorithm could build the SVM model in real time based on the automatic selection of test cases, and output the quantization model. In addition, the ARED method is compared with the SFE and FE method, under the bimodal Gaussian curve function, the bimodal surface function and the peaks function, the following conclusions are obtained.

1). The more complex the function form and the higher the dimension, the more test cases requires. In the test of the bimodal Gaussian curve function, the average number of cases is 12.7 times; in the test of the bimodal surface function, the average number of cycles is 28.4 times; and for the test of the peaks function, it is 35 times.

2). The ARED method can be applied to multi-factor experiments, and only the control parameters need to be adjusted. This paper gives an empirical combination of control parameters, p take , v take under the two dimensional sample set, as for three dimension, p take, v take.

3). Compared with the single factor experiment method (2D) and the factorial experiment (3D), the ARED method has greater precision and applicability. As the test dimension increases and the law of experimental complexity increases, the advantages of ARED method will be more obvious.

**Acknowledge**

This research did not receive any specific grant from any funding agencies in the public, commercial, or not-for-profit sectors.

**Support information**

Fig. S1 The selected test cases of ARED and SFE in bimodal Gaussian function

Fig. S2 The selected test cases of ARED and SFE in bimodal surface function

Fig. S3 The selected test cases of ARED and FE in peaks function.

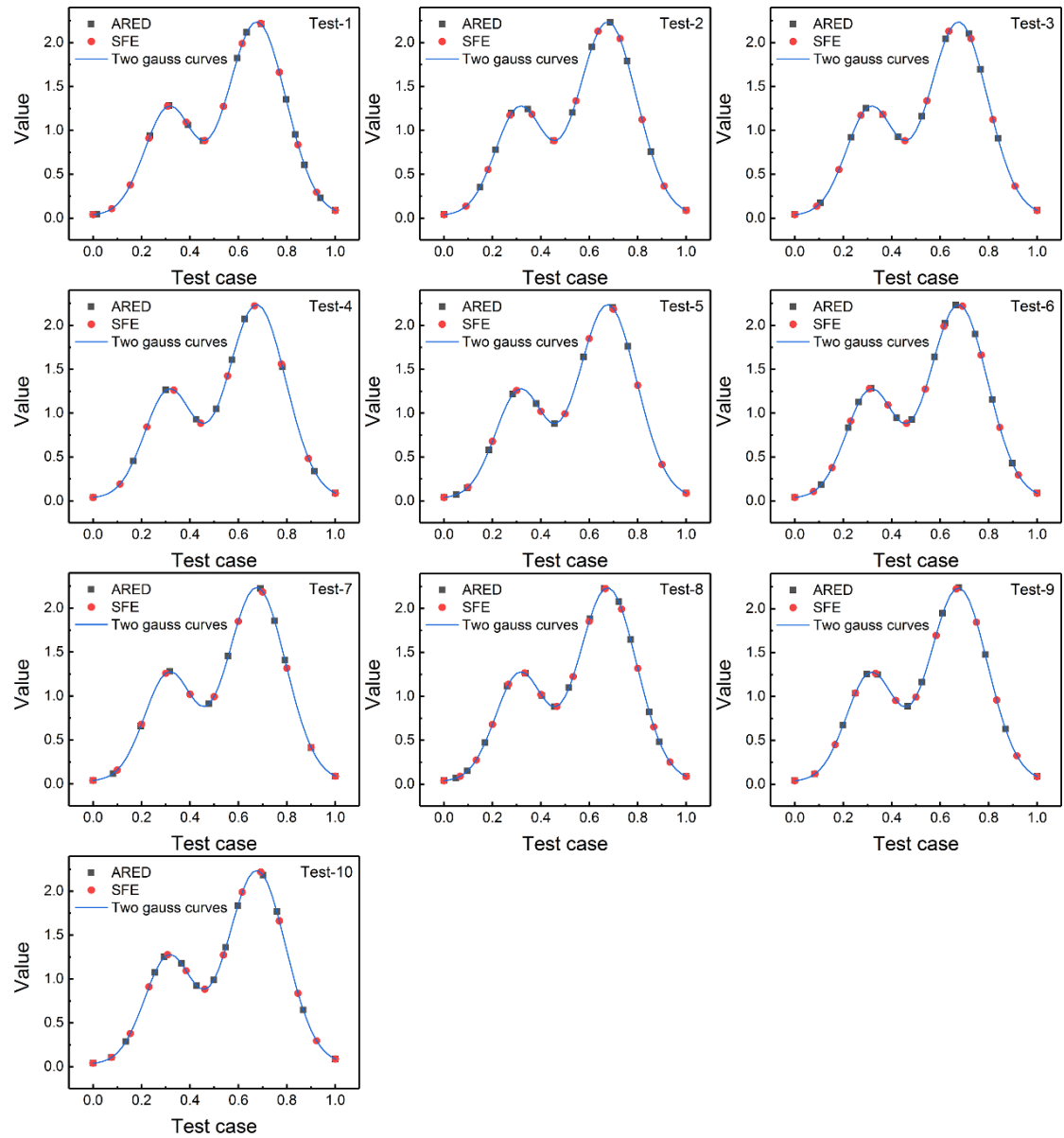

Fig. S1 The selected test cases of ARED and SFE in bimodal Gaussian function

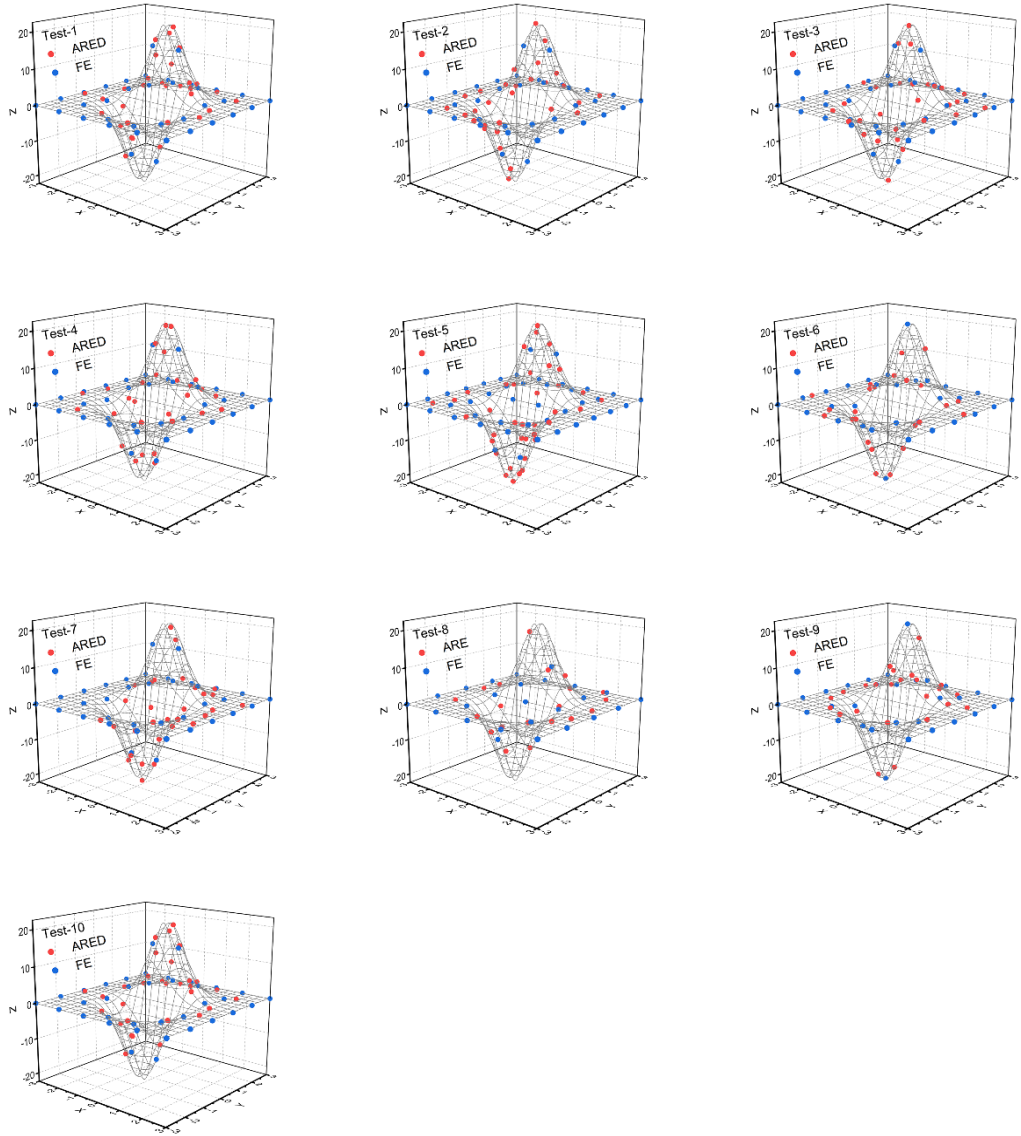

Fig. S2 The selected test cases of ARED and SFE in bimodal surface function

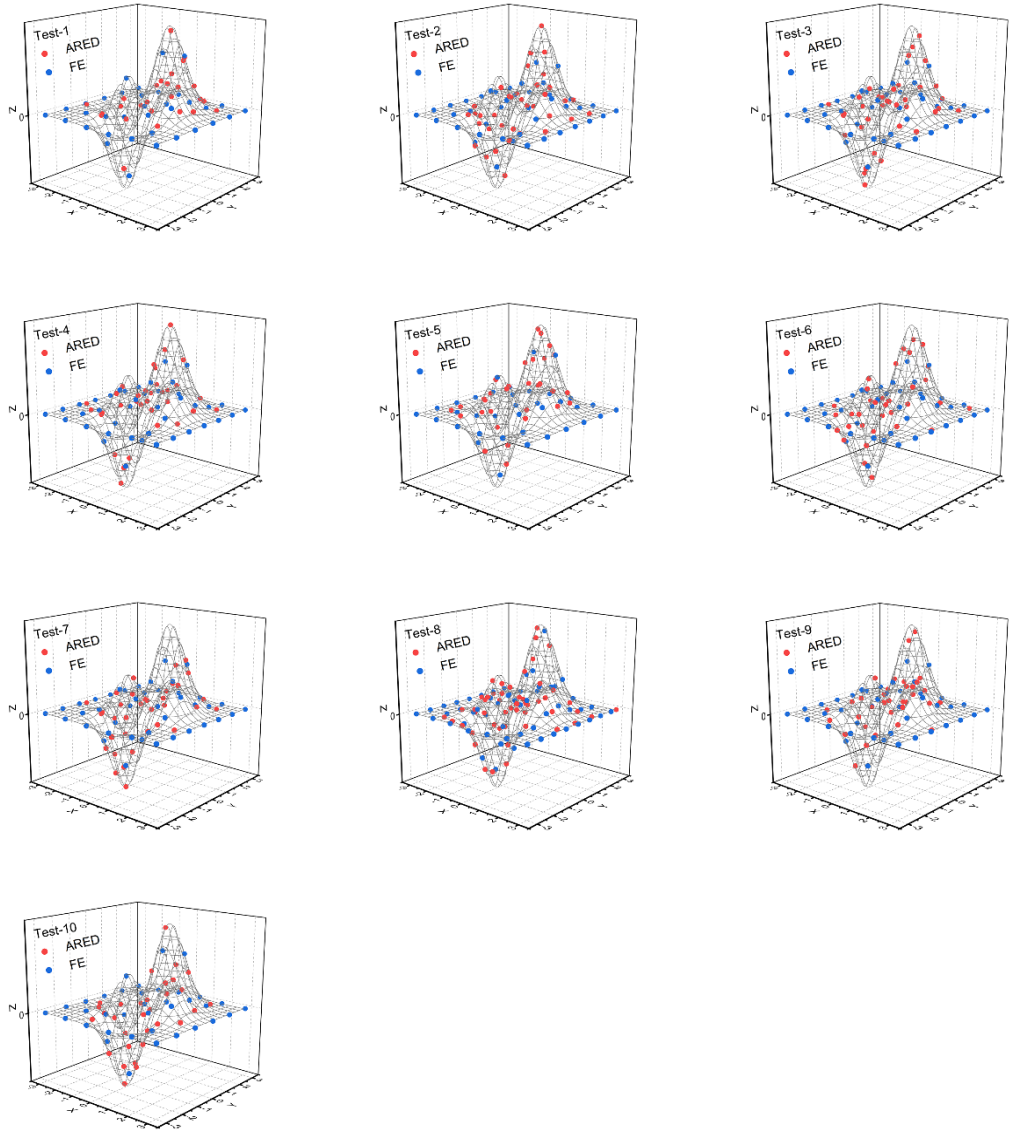

Fig. S3 The selected test cases of ARED and FE in peaks function.